\documentstyle[12pt,aas2pp4]{article}
\begin{document}
\def\today{\number\day\space\ifcase\month\or
  January\or February\or March\or April\or May\or June\or
  July\or August\or September\or October\or November\or December\fi
  \space\number\year}
\def\etal{{\it et al.\/}}
\def\ie{{\it i.e.\/}}
\def\kms{{$\,{\rm km\,s}^{-1}$}}
\font\smcap=cmcsc10
\def\hi{{\smcap H$\,$i}}
\def\hii{{\smcap H$\,$ii}}
\def\ha{H$\alpha$}
\title{Tests of the Tully-Fisher Relation II:  Scatter Using
Optical Rotation Curves$^1$}
\author{Somak Raychaudhury}
\affil{Inter-University Centre for Astronomy and Astrophysics, 
Pun\'e 411~007, India}
\affil{\tt somak@iucaa.ernet.in}
\medskip\medskip
\author{Kaspar von Braun and Gary M. Bernstein}
\affil{Astronomy Department, University of Michigan, 830 Dennison Building,
Ann Arbor, MI~48109, USA}
\affil{\tt kaspar, garyb@astro.lsa.umich.edu}
\medskip\medskip
\author{Puragra Guhathakurta}
\affil{UCO/Lick Observatory, University of California, Santa Cruz, CA 95064,
USA}
\affil{\tt raja@ucolick.org}

\altaffiltext{1}{Based on observations obtained with the 
Multiple Mirror Telescope, a joint facility of the Smithsonian
Institution and the University of Arizona.}

\begin{abstract}
We investigate the amount of scatter in the Tully-Fisher relation
(TFR) when using optical long-slit \ha\ rotation curves to determine
the velocity widths of spiral galaxies.  We study a sample of 25
galaxies in the Coma region of the sky which were shown in Bernstein
\etal\ \markcite{B1}
(1994) to exhibit an extraordinarily low scatter of 0.10~mag
RMS in the $I$ magnitude vs 21-cm width TFR.  Using the same $I$
magnitudes with new widths derived from high-quality \ha\ rotation
curves, we measure an RMS scatter of 0.14~mag in the TFR.  This
suggests that measurement errors and ``astrophysical errors'' (such as
non-circular gas motion) on the \ha\ velocity widths are below 6\%,
and optical widths are nearly as good for TFR studies as 21-cm widths.
The scatter and form of the TFR are found to be robust under choice of
velocity width-extraction algorithm, as long as the central portions
of the optical rotation curve are ignored and low-S/N points are not
weighted too heavily.  In this small sample there is no evidence that
rotation curve shapes vary systematically with rotation velocity, nor
that rotation curve shape can be used to reduce the scatter in the
TFR.

\end{abstract}

\keywords{cosmology: distance scale; galaxies: spiral; galaxies: fundamental
	parameters; galaxies: distances and redshifts}
\setcounter{footnote}{0}

\bigskip
\begin{center}
 Accepted for publication in the {\it Astronomical Journal}
\end{center}

\section{Introduction}

Peculiar velocity surveys of thousands of spiral galaxies are now
being conducted using the Tully-Fisher relation (TFR, Tully \& Fisher
1977) between the speed of rotation of a spiral and its absolute
magnitude.  The majority of these surveys have been conducted using
21-cm \hi\ rotation widths, but many now also make use of long-slit
optical \ha\ rotation curves to determine the rotation speed ({\it
e.g.\/} Matthewson, Ford, \& Buchhorn \markcite{M1} 1992, 
Giovanelli \etal\markcite{G1}\markcite{G2} 1997ab,
Willick \etal\ \markcite{W1} 1996).

In this paper we ask how the scatter in the TFR using optical data
compares to the 21-cm TFR scatter.  This comparison is facilitated by
the use of a sample of 25 galaxies in the Coma Supercluster region
which we have previously shown to have extraordinarily low scatter in
the $I$ magnitude vs 21-cm width TFR---only 0.10~mag RMS (Bernstein
\etal\ \markcite{B1} 1994, hereafter referred to as Paper I).  We have
obtained high-quality \ha\ rotation curves for all these galaxies,
have extracted velocity widths, and determine the TFR scatter using
optical widths.  Since the photometry and 21-cm data for these
galaxies (either by chance or for some not-yet-understood physical
reason) give a nearly perfect 21-cm TFR, we are able to perform a very
sensitive test for additional scatter induced in the TFR by a switch
from 21~cm to \ha\ linewidths.  The scatter in the optical TFR will
present an upper limit on the TFR scatter associated with the optical
widths, whether the cause be measurement error or some failure of the
rotation curves to reflect the ``true'' rotation width.

\section{Expectations and Previous Results}

{\it A priori}, 21-cm widths are expected to be superior to optical
widths in the context of TFR studies.  The detectable 21-cm emission
extends further out into the disk of the spiral than the \ha\
emission, and the gas motions are likely more regular and circular in
the outer disk than in the central sections.  The 21-cm velocity
should then better reflect the presumed underlying physical variable
of the TFR, galaxy mass, and thus produce lower TFR scatter.
Secondly, all \hi\ atoms in the galaxy contribute equally to the 21-cm
profile (dubbed ``atomic democracy'' by Schommer \etal\ \markcite{S1}
1993) so the data are not biased or restricted to areas of high
ultraviolet radiation.  Furthermore, 21-cm signals are not attenuated
by dust---though extinction of \ha\ photons is relatively small at
2--3 scale lengths into the disk where the most relevant dynamical
information lies.  Finally, it is easy to achieve $<10$\kms\
resolution in 21-cm observations.

Long-slit rotation curves have the advantage 
over 21-cm profiles of being a two-dimen\-sional
data set (spatial and velocity axes).  
The latter have no spatial information, so it
is not possible to distinguish thermal or non-circular motions from
true rotation.  On the long-slit rotation curve, random motions appear
as noise atop the true rotation and thus can be removed.  In choosing
an algorithm to extract a single rotation width from the rotation
curve, we may also restore some ``atomic democracy'' if we do not place
too much weight on \hii\ regions with high flux, even though such
regions might have very small formal errors on the recession velocity.
Furthermore the morphology of the rotation curve could provide extra
information to help reduce TFR scatter.

It is also possible to obtain 3-dimensional data (two spatial axes and one
velocity axis) using Fabry-Perot images of the \ha\ emission, as
demonstrated by Schommer \etal\ \markcite{S1} (1993).  In theory this extra
information should help detect dynamical irregularities even better
than long-slit data, and also allows determination of the galaxy
inclination and position angle independent of the photometric data.  By
simulating long-slit observations of their Fabry-Perot datacubes,
Schommer \etal\ demonstrate several potential problems for long-slit
measurements.  They conclude, however, that for highly inclined
galaxies differences between Fabry-Perot and long-slit estimates
of the rotation speed will be small.

Long-slit \ha\ measurements are perhaps the easiest to make in practice.
They certainly require less telescope time than Fabry-Perot
measurements, and can be extended to much more distant galaxies than
21-cm methods---especially in regions of the sky
which are inaccessible to the
Arecibo telescope.  Rotation curves of spirals at 7000\kms\ recession
velocity can be obtained in under an hour with 2-meter class
optical 
telescopes, but become difficult to observe with radio telescopes 
other than Arecibo.  Of course a practical disadvantage to long-slit
observations is that one must know the position angle of the galaxy
dynamical major axis
in advance (preferably to an accuracy of 2\arcdeg--3\arcdeg).  In our study
the photometric position angle is known to 1\arcdeg--2\arcdeg\ 
because we precede the \ha\ observations with very
deep $I$-band imaging, and we remove from our sample any galaxy in which
the outer isophotes twist by more than a few degrees.  It has been noted
(Franx \& de~Zeeuw \markcite{F1} 1992) 
that if spiral disks are not intrinsically round, then the photometric
major axis will not in general coincide with the dynamical major axis,
causing errors in long-slit observations.  Our optical TFR scatter may
be taken as a test of this non-circularity.

These practical considerations have led others to attempt TFR
measurements with optical rotation curves.  Courteau \markcite{C1}
(1992) conducted a TFR peculiar velocity survey with \ha\ rotation
widths, and investigated several algorithms for extracting a width $W$
from a rotation curve.  These are evaluated first by checking
variations on repeat measurements, and also by comparison with 21-cm
values.  As one might expect, algorithms which involve choosing a
maximum point on each arm of the rotation curve fare poorly because
they will usually extract the width from the noisiest bins.  The
lowest internal scatter in $W$ (10\kms) results from essentially
averaging all velocities over each arm of the rotation curve.
Courteau chose, however, to add all the spectra along the slit,
discarding all spatial information, and collapsing all the \ha\ flux
into a single velocity profile.  He then derived $W$ by means similar
to those used for 21-cm profiles.  While his spatially-averaged
optical widths gave higher internal errors (12--13\kms) than $W$
estimates which use the spatial information in the rotation curve, the
offset from 21-cm $W$'s is smaller, which made them better suited to
his purposes.

Other investigators have used \ha\ rotation curves to obtain $W$ but
do not explicitly compare various algorithms.  Mathewson, Ford, \&
Buchhorn \markcite{M1} (1992) subtract the minimum from the maximum of
the rotation curve to obtain $W$, and measure an internal scatter of
10~\kms.  As discussed below, it is likely that they have enforced
some unspecified minimum S/N on the rotation curve points.  Giovanelli
\etal\markcite{G1}\markcite{G2} (1997ab) 
reinterpret some of the Matthewson, Ford, and Buchhorn
rotation curves by defining $W$ to be the width at $R_{\rm opt}$ (the
83\%-light radius) from the galaxy center.  This requires
extrapolation for some rotation curves which do not reach this deep
into the disk (3.2 scale lengths).

Vogt \markcite{V1} (1994) extracted widths from rotation curves using
a variant of the velocity-profile method: she measured the mean
velocity in each spatial bin along the rotation curve, ranked these
velocities in order, and defined the rotation width to be the
difference between the 10th and 90th percentile points of the velocity
distribution.  This should encourage more atomic democracy than the
Courteau algorithm, because it weights all parts of the rotation curve
equally rather than by flux.

\section{Measurements of Rotation Curves}

\subsection{The Sample}
The Coma Supercluster galaxy sample and the collection of photometric
data are described in Paper I.  Briefly, we select those galaxies
within a few degrees of the Coma cluster core, with recession
velocities in the range 5000--8000\kms, for which 21-cm profiles could
be found in the literature.  Deep $I$-band surface photometry was
obtained for each galaxy.  We discarded galaxies for which the
ellipticity varied by more than 0.03 across the outer isophotes or for
which the position angle twisted by more than $\approx3$\arcdeg---{\it
e.g.\/}, those with tidal tails, those for which the isophotal shape
was dominated by the spiral arms, or those with morphological
peculiarities.  For these discarded galaxies ($\sim$15\% of the full
sample), we have little confidence that the isophotal shape accurately
reflects the disk inclination, making it difficult to derive the
rotation speed from line-of-sight velocities along the photometric
major axis.  We also removed from the sample those galaxies ($\sim$6\%
of the full sample) for which the 21-cm profile did not have
sufficiently steep sides to obtain a reliable width.

In Paper~I we demonstrate that few (if any) of the galaxies in our
sample are likely to be members of the Coma cluster itself, and we
henceforth assume that their distances are proportional to their
redshifts.

\subsection{Observations}

Each of the galaxies was observed in March 1993 or June 1993 using the
Red Channel spectrograph on the Multiple Mirror Telescope (effective
aperture of 4.5m).  A 1200 lines~mm$^{-1}$ grating blazed at 575~nm
gives a resolution of $\sim$0.21~nm per pixel near the \ha\ line.  
In addition, the
thinned $800\times1200$ CCD was binned to 
0\farcs6 pixels in the spatial direction.  The
slit dimensions were $1\farcs25\times180\arcsec$, with the width
chosen to admit as much light as possible without significantly
degrading the spectral resolution.  Most of the galaxies were observed
for a single 1200~s exposure, with some of the fainter galaxies
observed for two 1200~s exposures.  This is significantly deeper than
most existing TFR rotation curve studies---we wish to investigate the
``intrinsic'' errors in the TFR using rotation curves rather than have
the scatter be dominated by errors from photon noise.  Position angles
for the galaxies were chosen to match the outer isophotes of the
$I$-band images.  Galaxy nuclei were centered on the slit using the
MMT guiding camera.

The galaxy observations were interspersed with short wavelength calibration
exposures of a He-Ne-Ar arc lamp, and quartz lamp flat-field exposures.
Quartz lamp spectra were also obtained through a ``decker'' slit consisting
of a row of small holes.  Long-slit spectra of the twilight sky were also 
obtained.

\subsection{Extraction of Rotation Curves}
Each galaxy exposure was flat-fielded using the quartz lamp exposure
adjacent in the observing sequence, 
and a wavelength solution for the entire CCD was
derived from the adjacent He-Ne-Ar arc lamp exposure.  Residuals to the
wavelength fit are typically 0.005~nm.  Twilight exposures were used to
compensate for slight 
differences (dependent on the position of the slit) between lamp
throughput and sky throughput caused by small differences between the
pupil illumination by the lamp and by the sky.
Finally, the decker slit quartz lamp exposures were
used to map the distortion of the spatial coordinate on the
chip.  Thus for each exposure we produce a map from pixel coordinates
$(x,y)$ to $(s,\lambda)$ space (slit position vs wavelength).

For a given pixel corresponding to $(s_i,\lambda_j)$, we determined the sky
background as follows:  we determined the flux at $\lambda_j$ for each
other row ($s$ value) in the image.  Note that since the spectrum is
tilted relative to pixel coordinates, this may require interpolation.  
We then fit a linear function of $s$ to the derived sky background intensity
vector, omitting of course the central parts of the slit over which
the galaxy emission can be seen on the CCD image.

Cosmic rays were identified by eye, and the IRAF task IMEDIT was used
to linearly interpolate over pixels affected by cosmic rays.  Each
pixel was then assigned to a bin according to its spatial coordinate
$s$.  Bin widths were 3 to 10 pixels (1\farcs8--6\arcsec) depending
upon the S/N of the emission lines.  Within each spatial bin, the
intensities in each spectral pixel were fit to a model function of
wavelength $\lambda$ which includes continuum radiation and gaussian
profiles for the \ha\ and the two {[\smcap N$\,$ii]} lines
(undersampling of the spectral line profile is accounted for).  Each
spectral line was allowed to have an independent intrinsic width
(which was then convolved with the instrumental width) and flux, but
the three lines were constrained to have the same recession velocity.
Thus for a linear background, the spectral model has 2 parameters for
continuum, 6 line strength/width parameters, and one velocity
parameter.  This model was fit to an $\approx80$-pixel ($\sim$17~nm)
segment of the spectrum spanning the \ha\ and {[\smcap N$\,$ii]}
lines, leaving $\approx70$ DOF for the fit.  Chi-squared values were
typically 70-100, except near the galaxy nucleus, where the continuum
has high enough S/N to make the linear approximation inadequate.  Note
that we {\it do not rebin or interpolate} the spectra.

The fitting process resulted in an estimate of recession velocity $v_i$
and its uncertainty $\sigma_{v_i}$ for each bin, with the bins centered
at positions  $s_i$ along the slit.  We only used bins in which the
\ha\ emission was detected at $\rm S/N>3.5$, which leads to velocity
uncertainties smaller than 12\kms\ even in the noisiest parts of the rotation
curves.  Rotation curves for the 25 galaxies are shown in Figure~1.

\begin{figure*}[t]
\figurenum{1a}
\epsscale{1.5}
\plotone{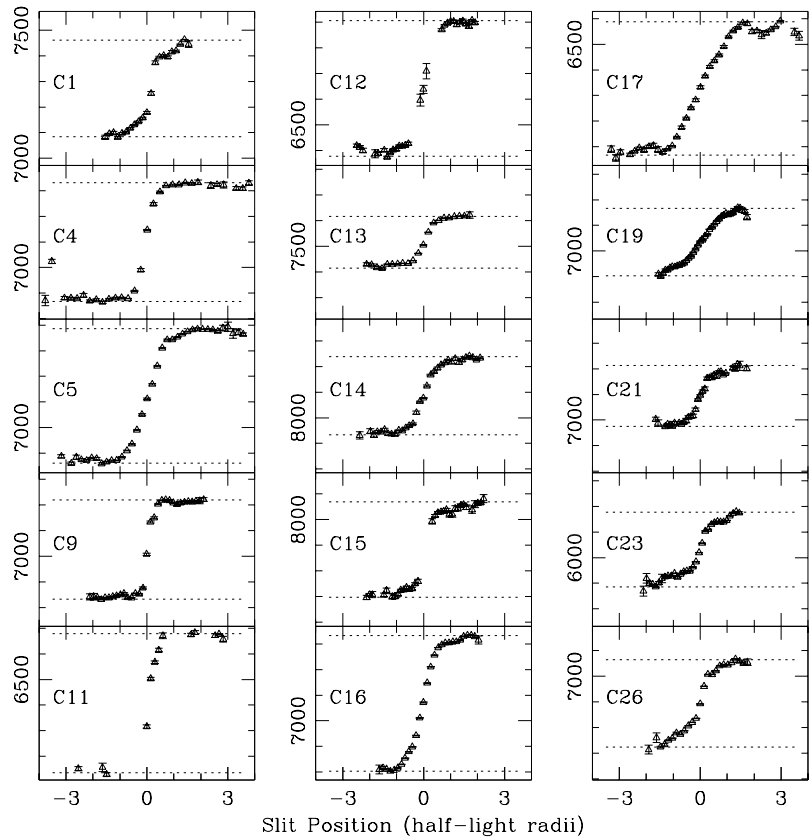}
\caption{Rotation curves for the 25 galaxies in the sample.  Bins
with S/N below 3.5 are omitted.  The $x$ axis is in units of each
galaxy's half-light radius $r_{1/2}$, and the $y$ axis is recession
velocity in \kms.  All galaxies are on a common scale with each
vertical tick being 100 \kms.  Dashed lines are the velocities of the
two arms as determined by the Probable Min-Max method.}
\end{figure*}

\begin{figure*}[t]
\figurenum{1b}
\plotone{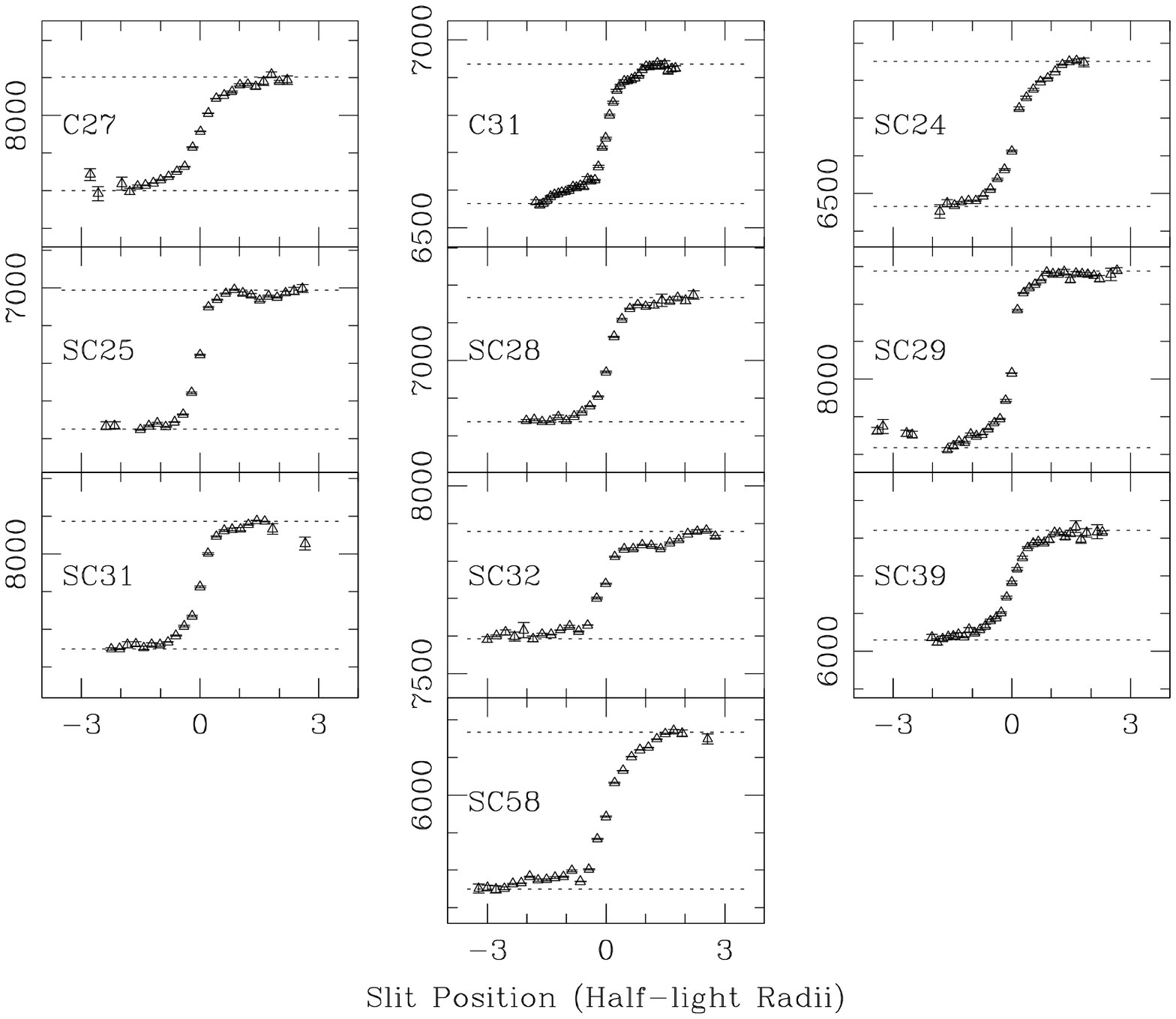}
\caption{Rotation curves for the 25 galaxies in the sample (continued).
}
\epsscale{0.95}
\end{figure*}

\section{Velocity Width Algorithms}

We must extract a single width $W$ from the rotation curves in
order to produce a TFR.  Desirable qualities for the $W$ 
measurement algorithm are:
\begin{enumerate}
\item The resultant $W$ should be robust under changes in binning of
	the rotation curve data (or changes in spatial resolution).
\item $W$ should also vary little with the S/N level of the observations.
\item $W$ should have minimum variance for repeat observations at
	a given S/N level of the spectrum.
\item The algorithm should try to promote ``atomic democracy'' (see \S2).
\item The algorithm should not make {\it a priori} assumptions about
	the shape of the rotation curve.
\item Most importantly, it should minimize the scatter in the TFR.
\end{enumerate}

The first three of these criteria lead us to design algorithms
that use as much of the rotation curve as possible in some form
of averaging or fitting process---though criterion (5) leads us to
avoid fitting functional forms to the curves.  Choosing the maximum
and minimum velocity of the rotation curves (and subtracting to
get $W$) is a very poor method, because the final $W$ is
determined by the noisiest bins.  The derived $W$ will
have a strong tendency to increase
as S/N worsens or spatial bin size is decreased.
We describe here several width-determination algorithms which we
apply to the data in order to see which performs best under
criterion (6), minimal TFR scatter.

\subsection{Min-Max Method}
This is the ``straw man'' method upon which we hope to improve.  
The width $W$ is
simply $v_{\rm max}-v_{\rm min}$, where $v_{\rm max}$ and $v_{\rm min}$
are the maximum and minimum of the $v_i$.  This indeed leads to quite
undesirable results, with a TFR scatter of 0.4~mag.  The noise in
the estimate of $W$ is
much abated if we consider only those $v_i$ for which the \ha\ line has
$\rm S/N>3.5$ and $\sigma_{v_i}<12$\kms; we will henceforth retain this
restriction. Some form of this algorithm is used by Matthewson, Ford,
\& Buchhorn \markcite{M1} (1992), and Courteau \markcite{C1} (1992) also experiments with variants
of this method.  The ``percentile'' method of Vogt \markcite{V1} (1994)---choosing the
width to be the difference between the 10th and 90th percentile points
of the velocity histogram---is similar to the min-max method, but with
much reduced sensitivity to the noise level.  Vogt also reduces
sensitivity to noise by using only points with small uncertainties.

\subsection{Weighted Mean Method: Entire Side}
The width $W$ is here defined to be
$|v_{\rm left} - v_{\rm right}|$, where the left and right side
velocities are the weighted averages of the entire side of the rotation
curve.  The center is located at the maximum of the continuum flux.
The velocity data points
are weighted by $\sigma_{v_i}^{-2}$ as usual; this might
subvert ``atomic democracy'' by weighting too heavily those bright
\hii\ regions with high flux and thus low $\sigma_{v_i}$.  We could
encourage atomic democracy by replacing $\sigma_v$ by some minimum
value ($\approx1$~\kms) if the flux is high enough to make the formal
error smaller.  For our sample, however, we find that this does not make
a significant difference for any of the weighted mean algorithms
considered.

\subsection{Weighted Mean Method:  Outer Segments}
This algorithm is the same as the previous algorithm, except that the
inner part of the rotation curve is not used in calculating the
weighted means.  We define $s=0$ to be the center of the rotation curve,
and the extent $S$ of each side of the rotation curve is
defined by outermost $s_i$ which satisfies the S/N limit.  We then
perform a weighted mean on each side for all points with $S/2 < s_i < S$.
Courteau \markcite{C1} (1992)
obtains his best internal consistency with a similar algorithm.

\subsection{Weighted Mean Method:  Half-light Radii}
We improve upon the ``outer segments'' algorithm by averaging the
$v_i$ (taking a weighted mean) over an interval in $s$ located a fixed
number of half-light radii from the center of the galaxy.  Half-light
radii $r_{1/2}$ (actually half-light semi-major axes) are determined
from the $I$-band photometry and thus, unlike the ``lengths'' of the
rotation curves above, are independent of galaxy distance or rotation
curve S/N.  We define the velocity of each arm to be the weighted mean
over the range $[r_{1/2},2r_{1/2}]$.  We also calculate widths using
the range $[r_{1/2},4r_{1/2}]$.  For comparison we note that an
exponential disk galaxy has $r_{1/2}$ equal to 1.68 scale lengths, and
also has $r_{1/2}=0.52R_{\rm opt}$, where $R_{\rm opt}$ is the
83\%-light radius defined by Persic, Salucci, and Stel \markcite{P2}
(1996, hereafter referred to as PSS96).

\subsection{Fixed-Point Method}
We measure the velocity of each arm of the rotation curve at a fixed
number of half-light radii from the center.  The $v_i$ are
interpolated to the appropriate point; only points with $\rm S/N>3.5$
are used for interpolation.  We find that the lowest scatter is
obtained by choosing the fixed point at $1.3r_{1/2}$ from the center,
though the TFR scatter depends only weakly upon the exact choice as
long as the fixed point is beyond $r_{1/2}$.  Giovanelli \etal\markcite{G1} 
\markcite{G2} (1997b), for example, choose to measure
$W$ at $R_{\rm opt}=1.9r_{1/2}$.  We choose not to use such a large
radius because for a few of our galaxies (and those of Matthewson,
Ford, and Buchhorn \markcite{M1} [1992] which Giovanelli \etal\ use),
this requires extrapolation past the high-S/N region of the rotation
curve.

\subsection{Probable Min-Max Method}
The above algorithms determine some average rotation velocity,
or velocity on a robustly determined section of the rotation curve.
It may be that the TFR is tighter when $W$ is a measure of the
maximum rotation speed of the galaxy instead of some average
rotation speed.  Unfortunately a simple maximum of the $v_i$ is
not robust (Sec.~4.1).  Here we define a fairly robust means of quantifying
the extremes of the rotation curve.  We define $v_{\rm max}$ as
that which satisfies
\begin{equation}
\prod_i P(v_{\rm max}>v_i) = 0.1;
\end{equation}
where the probabilities $P$ are given by the standard error function
if we assume that $v_i$ has a gaussian distribution about its
measured value, with dispersion $\sigma_{v_i}$.  
In effect we
ask the following question:  At what $v_{\rm max}$ is it likely
(at the 90\% level) that {\it some part} of the rotation
curve exceeds $v_{\rm max}$?  A single high point on the rotation
curve with a large measurement uncertainty does not
unduly influence this estimator,
and it is fairly insensitive to binning as well.  We
define an analogous $v_{\rm min}$ on the other side of the rotation curve
and define $W$ as the difference $v_{\rm max}-v_{\rm min}$.

\section{Tully-Fisher Scatter}
For each of the above algorithms for measuring $W$ 
we evaluate the slope and scatter of
the resultant TFR.  As explained in Paper I, we fit to the TFR data
a model of the form
\begin{eqnarray}
\label{tfeq}
\lefteqn{
I_{\rm tot} - 5\log(z/0.0233)}  \nonumber\\
& - (1-\Omega/2+k_I)(z-0.0233) = \nonumber\\
& I_0 + m \log( { {W_0} \over {400\,{\rm km\,s}^{-1}} }) + a_I e.
\end{eqnarray}
The left-hand side is the apparent total magnitude corrected to a
standard redshift of $z=0.0233$, or a recession velocity of 7000\kms,
assuming all galaxies to be in free Hubble expansion.
Recession velocities are taken as the mean of the two arms of the
rotation curve, and in general agree very well with the radio data.
The $k$-correction factor $k_I$ is fixed at 0.6.  On the right-hand
side, $e$ is the galaxy ellipticity as determined from the $I$-band
surface photometry, and $W_0$ is the measured $W$ corrected to edge-on by
the sine of the inclination angle, as described in Paper I.  Note 
no correction is made for turbulent motion.  The free parameters in the fit
are $I_0$, $m$, and the inclination correction coefficient $a_I$.

Table~1 lists the reduced photometric data for the 25 galaxies (raw
data may be found in Paper I) along with the rotation and recession
velocities determined from the optical and radio data.  In Table~2 we
show the results of the TFR fits using widths derived by the above
algorithms.  Figure~2 shows an example Tully-Fisher plot produced with
the Probable Min-Max widths.

\begin{figure*}[t]
\figurenum{2}
\plotone{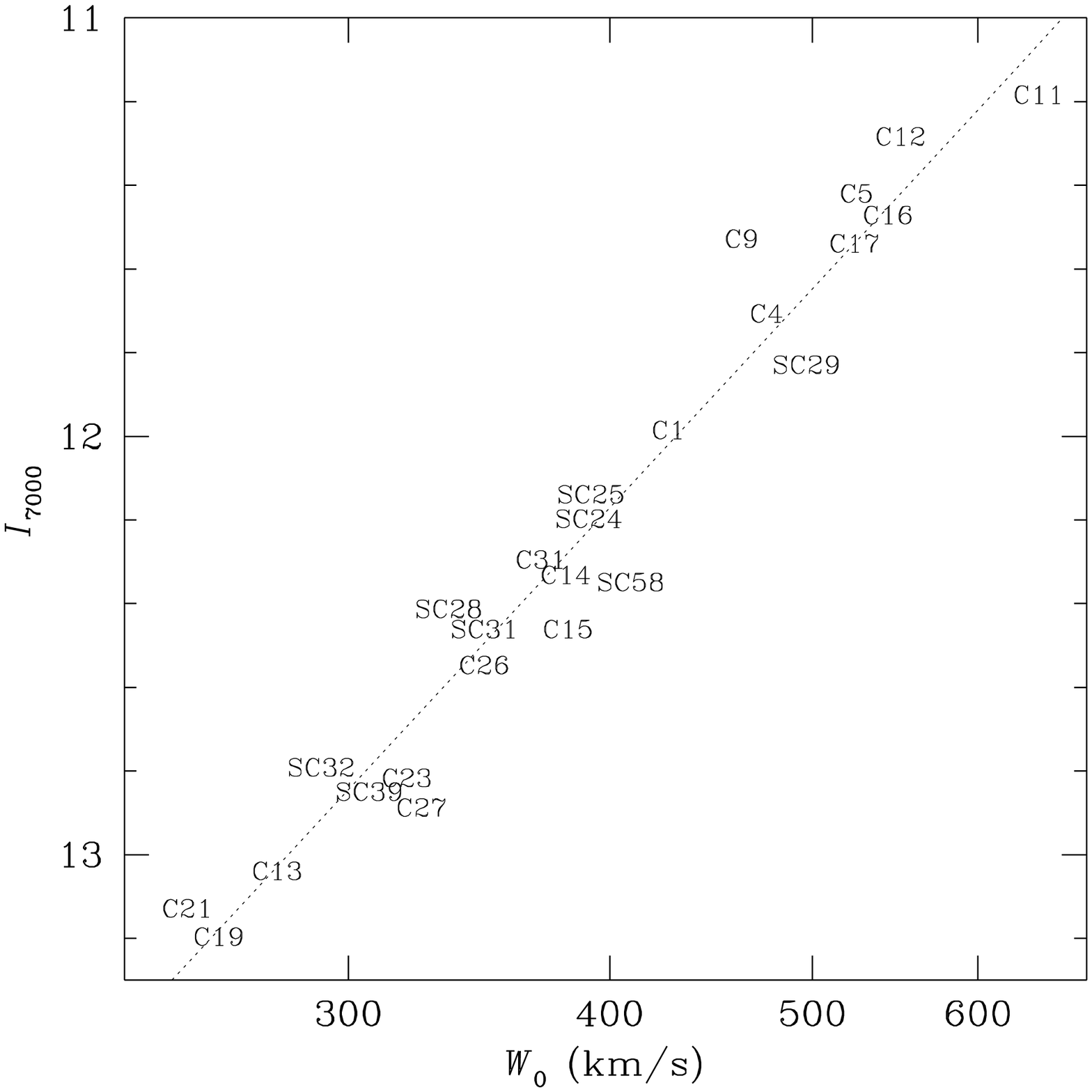}
\caption{Tully-Fisher relation for the 25 galaxies using widths
derived from \ha\ rotation curves with the ``Probable Min-Max'' method
described in the text.  Widths have been corrected to edge-on as
described in Paper I, and $I$-band total magnitudes have been corrected
to a common distance of 7000~\kms, and corrected for inclination using
the value of $a_I$ shown in Table~2.  The RMS deviation from the best-fit
dashed line is 0.14~mag.
}
\end{figure*}

The TFR scatter based on optical $W$ data is found to be quite robust.
Nearly all methods for determining $W$ give an RMS TFR scatter of
$\approx0.14$~mag.  Even the simple Min-Max method works well if we
consider only points with $\rm S/N>3.5$ (see \S4.1).  The only
algorithm noticeably less accurate than the others is the Entire Side
Weighted Mean, which leads us to believe that one should avoid the
central portions of the rotation curve in deriving $W$.

\subsection{Intercomparison of Widths}
Although it is not our goal to produce optical widths which best
match the 21-cm widths, we do note that there are systematic
differences among the different optical $W$ estimators, and between
optical and 21-cm estimates of $W$.  As expected, those algorithms
which attempt to measure the extremes of the rotation curve tend to
give larger $W$ values than those which attempt to measure the mean
over some segment of the rotation curve.  This can be seen by
comparing columns (8) and (9) in Table~1.  In galaxies which are more
face-on or have smaller rotation speeds, we expect the turbulent
motions to be a larger fraction of the observed velocities.  If
different width measurement methods have varying susceptibility to
turbulent motions, this might be manifested as changes in the slope
$m$ and/or inclination correction $a_I$ in the derived TFRs.
Note, though, that the values of $m$ and $a_I$ in Table~2 are consistent among
various algorithms, and are consistent with the values obtained using
21-cm widths. 

\begin{figure}[t]
\figurenum{3}
\plotone{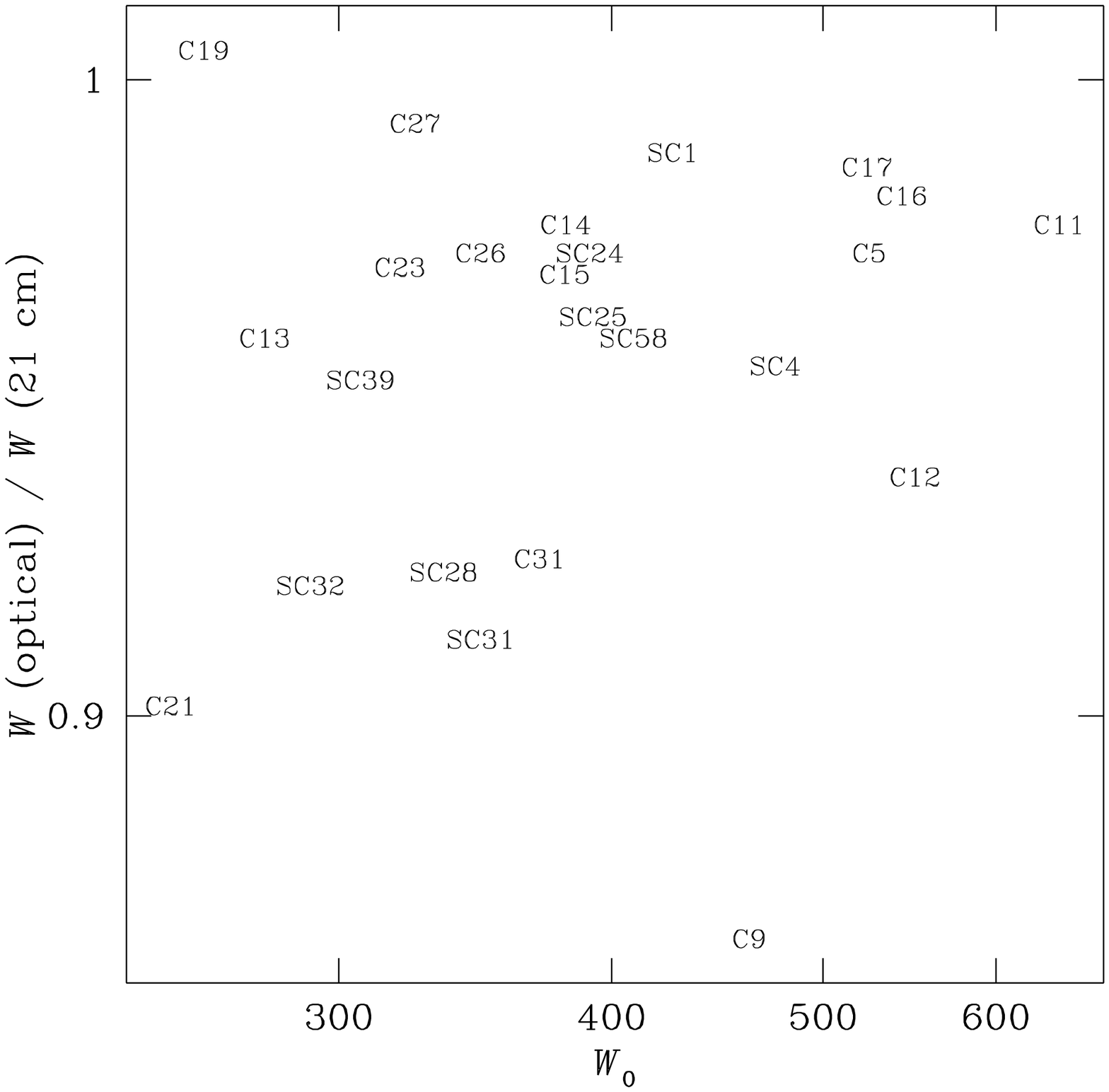}
\caption{Ratio of the velocity widths
derived from \ha\ rotation curves (using the ``Probable Min-Max''
method) to the 21-cm velocity widths from Paper I, plotted vs the
inclination-corrected optical width.  Optical widths are on average a fixed
fraction (96\%) of the radio widths.  There is no detectable
correlation between this ratio and the magnitude of the width, nor
with disk inclination (latter not illustrated here).
}
\end{figure}

Figure~3 plots the ratio of Probable Min-Max width to
21-cm width for all 25 galaxies as a function of inclination-corrected
width $W_0$.  No trend with $W_0$ is present; the optical widths are, on
average, 0.96 times the 21-cm widths, with no detectable dependence
upon $W_0$ or upon inclination.  For this sample of relatively large,
edge-on galaxies, the effect of turbulent velocities upon the 21-cm
profiles (or perhaps the effect of extinction on the \ha\ rotation
curves) is small, or at least constant across the full sample.

This 25-galaxy sample also gives us little guidance on how to best
treat rising rotation curves.  In Figure~1 we see that several of the
galaxies have rotation curves which are still rising at the ends,
particularly C19.  Should we take the ``true'' width to be the largest
observed width (using some form of Min-Max method), or something
larger (an extrapolation), or something smaller (one of the weighted
mean algorithms) to get the lowest TFR scatter?  Which of the
algorithms produces the best agreement with 21-cm widths?  The
near-equality of TFR scatter for all our algorithms means that we
cannot offer advice on this question.  The location of C19 near the
top of Figure~3 suggests that the probable Min-Max $W$ estimate of the
rising rotation curve {\it overestimates\/} the rotation speed
relative to 21-cm, so an extrapolation of the rising curve would make
agreement worse.  Even so, C19 is not an outlier from the TFR in
Figure~2.  The one outlier in the optical vs 21-cm comparison, C9, is
indeed the furthest outlier in the optical TFR (but is not an outlier
in the 21-cm TFR).  In fact most of the extra scatter in the optical
TFR over the 21-cm scatter is attributable to C9.  The rotation curve
of C9 is, however, beautifully flat and square, so a change of
algorithm would not lessen its departure from the optical
TFR. Furthermore, C9 has no unusual features which would justify
excluding it from our sample.

\section{Rotation Curve Shapes and Third Parameters}
In principle the rotation curve provides more information than a
simple width $W$---it has been suggested that some quanitification of
the {\it shape} of the rotation curve could be used to decrease the
TFR scatter (Persic \& Salucci \markcite{P1} 1990).  Large homogeneous
samples of rotation curves show that fast-rotating galaxies tend to
have rotation curves which rise more steeply in the center than
slowly-rotating galaxies, with the fastest-rotating galaxies actually
having a peak in the rotation curve (PSS96).

We wish to test whether, in our limited sample, there is a trend of rotation
curve shape vs $W_0$, or whether rotation curve shape can be
used as a third parameter to reduce the TFR scatter.  For each galaxy we
compare the area under both arms of the rotation curve to the area under a
perfectly flat rotation curve:
\begin{equation}
f \equiv { \int_{-s_{\rm lim}}^{+s_{\rm lim}}\,|v(s)-v_0|\,ds \over
	  2v_{\rm rot}s_{\rm lim}} .
\end{equation}
Here $s$ is the distance from the center of the galaxy, $v_0$ is the
recession velocity at the center, and $\pm s_{\rm lim}$ are the outer bounds
of the integration.  The rotation velocity $v_{\rm rot}$ is one-half
the width $W$, so the value of $f$ depends upon our choice of width
algorithm. If we choose $v_{\rm rot} = |v(+s_{\rm lim})-v(-s_{\rm lim})|/2$,
a solid-body rotator, would have $f=0.5$,
while a perfectly flat rotation curve would have $f=1$.

Table~1 contains the calculated value of shape parameter $f$ for each
galaxy. We determine $v_{\rm rot}$ using the Weighted Mean method over
the range 1--4$r_{1/2}$, and set $s_{\rm lim}=1.3r_{1/2}$.  
For comparison with the study of PSS96, we note that
they define $R_{\rm opt}$ to be the radius enclosing 83\% of the light
of a galaxy. For an exponential disk with scale length $R_{\rm scale}$,
$R_{\rm opt}=3.2 R_{\rm scale}$, while $r_{1/2}=1.68 R_{\rm scale}$,
so $R_{\rm opt}=1.9r_{1/2}$.  Thus, $f$
should be closely related to the ``inner slope'' defined in PSS96, or
to the rotation curve slopes defined by other investigators ({\it
cf.\/} Vogt \markcite{V1} 1994 and references therein).

\begin{figure}[t]
\figurenum{4}
\plotone{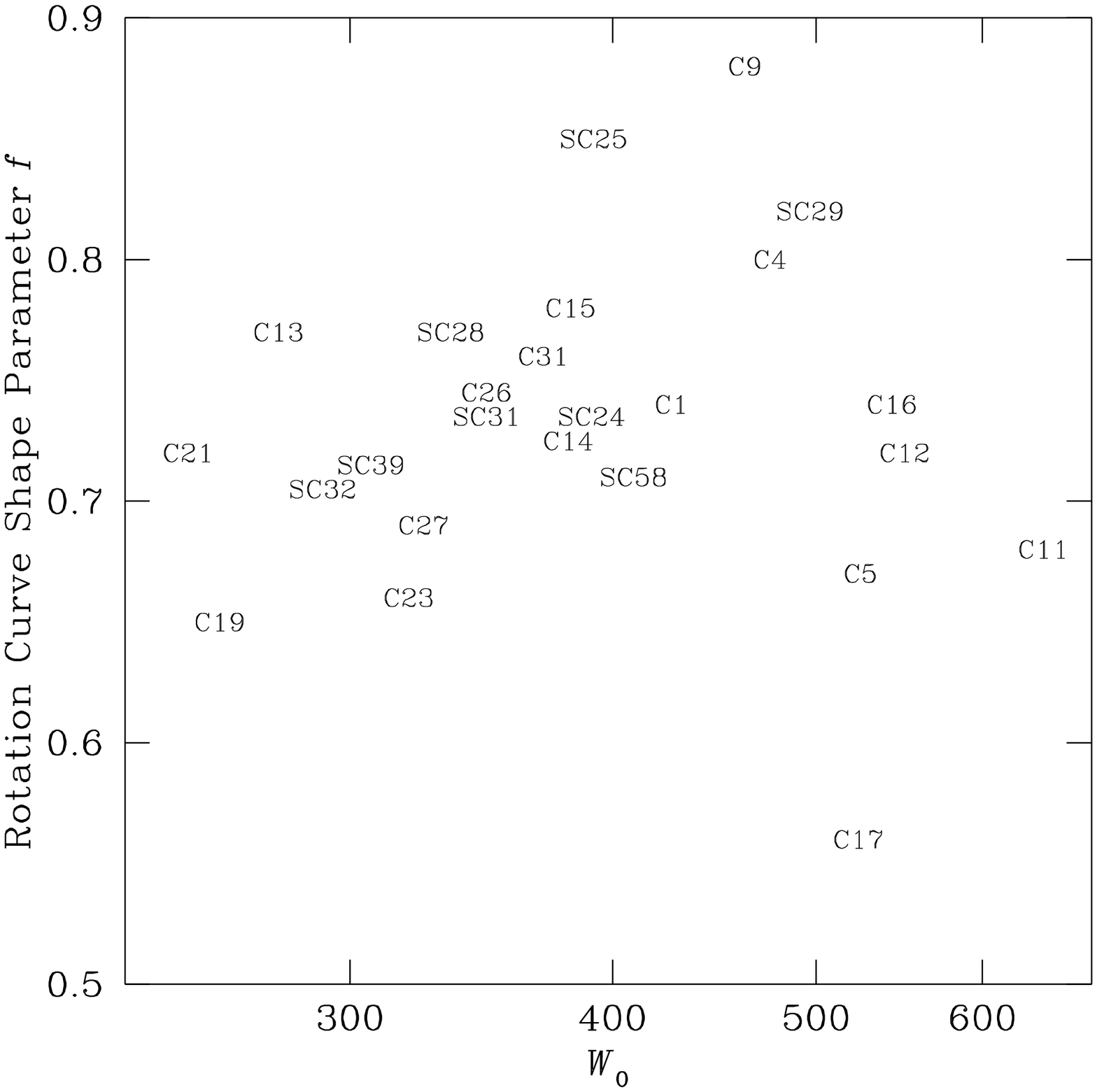}
\caption{Rotation curve shape vs rotation speed for the 25 galaxies.
The rotation curve shape, parameterized by the quantity $f$ defined
in Equation~(2), shows no significant correlation with inclination-corrected
rotation width $W_0$ on a galaxy-by-galaxy basis.  A galaxy with solid-body
rotation curve inside $1.3r_{1/2}$ would have $f=0.5$, while a step-function
rotation curve would have $f=1$.
}
\end{figure}

In Figure~4 we plot $f$ vs $W_0$ for the 25 galaxies.  There is no
significant correlation, indicating that, at least in this sample,
any trend of rotation curve shape vs amplitude is too weak to be
detectable in individual rotation curves.  This is, however, still
consistent with PSS96, because:  (1) we are looking at only a 2~mag
range in $M_I$, and the PSS96 ``universal'' rotation curves do not
change drastically over such a range, and (2) even in the PSS96 data,
the shape vs $W$ correlation is clearest only after co-adding many
rotation curves, and would not likely be detectable for 25 galaxies.
Note that galaxies C11 and C17 have small $f$ values although
substantial parts of their rotation curves are flat.  This is because
the flat parts lie outside our integration range of $\pm1.3r_{1/2}$;
a broader integration range would extend into regions of poor S/N for
some of our rotation curves and thus require undesirable extrapolation.

We also find that adding $f$ as a parameter to the TFR fit in
Equation~\ref{tfeq} does not reduce the resultant scatter.  Thus 
knowledge of the
steepness of the rotation curve does not, in this sample, seem to
improve our ability to predict absolute $I$ magnitude.

\section{Ramifications}
We find that the RMS scatter in the TFR using widths from long-slit
\ha\ rotation curves is $\approx0.14$~mag in this sample.  The TFR
slope and scatter are robust under choice of algorithm for extracting
the width $W$ from the rotation curve, as long as we ignore the central
section of the rotation curve and do not try to consider single points
with large uncertainties ($\rm S/N<3.5$ or $\sigma_v>12$\kms).  Using
$F$-test statistics we find that the optical scatter is greater than
the 0.10~mag scatter in the 21-cm TFR for these galaxies at 95\% 
confidence level.
If we ascribe {\it all} of the 0.14~mag scatter in the 
optical-$W$-based TFR to errors in
the $W$'s, the TFR slope of $m\approx-5$ implies an RMS
error of 6\% in the typical $W$ measurement.  Repeat measurements of a few
galaxy rotation curves have been made, some with slit position angle
intentionally
placed 10\arcdeg--20\arcdeg\ away from the major axis, and the resultant $W$ 
values are found to vary by 5\% or less; we believe that measurement error in
$W$ is not the dominant source of scatter in the optical TFR.
A fit to the inverse
Tully-Fisher relation gives nearly identical results because the
scatter is quite small.

``Errors'' in the $W$ values can take two forms.  First, there are
measurement errors, which would cause the measured value of $W$ to
differ each time we observed the galaxy.  Such errors might include
those due to photon statistics or poorly chosen slit position angles.
Second, there could be what we will term ``astrophysical''
errors---reasons why the \ha\ widths do not properly measure whatever
physical quantity is the underlying basis of the TFR.  Examples of
such cases might be velocities of \hii\ regions which depart from the
actual rotation velocity of the galaxy, or a poor choice of 
algorithm for measuring $W$
so that we do not measure the relevant rotation speeds
properly.  Of course since we do not actually {\it know\/} what the
underlying physical basis of the TFR is in any detail, we cannot
address these possible errors.

The scatter in our 21-cm TFR for these 25 galaxies is extraordinarily
low; most other TFR surveys find RMS variations of 0.3--0.4~mag (Paper
I).  This could be low by a statistical fluke: our draw from the pool
of either measurement errors or astrophysical errors ({\it i.e.\/}
choice of galaxies) has been lucky.  In either case, the low scatter
found with \ha\ widths means that any {\it additional} errors caused
by the use of optical widths is small (6\% or less in $W$).  Thus for
practical purposes, we conclude that {\it high-quality \ha\ rotation
curves may be used in place of 21-cm data for TFR studies, with no
practical loss in precision}.  This is also evident in the study of
Willick \etal\ \markcite{W1} (1996), who intercompare various TFR
surveys.  They derive RMS scatters of 0.40, 0.38, and 0.47~mag for
three surveys which use 21-cm widths, and 0.38 and 0.43~mag for two
surveys based primarily on \ha\ widths.

The scatter in the 25-galaxy sample could be very low because of some
as-yet-undiscovered physical commonality that is not present in the
larger TFR surveys (as the lower slope of our TFR might also suggest).
In this case we could also conclude that the measurement errors on the
optical $W$ values are small, a result which is probably extendable to
larger surveys.  If these 25 spirals differ physically from the
overall spiral galaxy
population, then our results on the robustness of the $W$ estimators
and on the insignificance of rotation curve shape may not be applicable
to the general spiral population.

We note also that the TFR slope and the inclination correction
coefficient $a_I$ are robust, with all optical $W$ algorithms giving
values consistent with the 21-cm values.  The constancy of the TFR slope
implies that \ha\ widths are in fixed proportion to the 21-cm widths
over the range in galaxy sizes explored here.  Note that our sample
consists primarily of large spirals.  The consistency of $a_I$ implies
that none of the algorithms is excessively subject to inclination
effects.  This means, for example, that the measured \ha\ widths are not
significantly changed relative to the ``true'' widths by the effects of
extinction in nearly edge-on
galaxies.  Since we made no corrections for turbulent velocities in
either the 21-cm or the \ha\ data, we also infer that such corrections
are not too important for the high-inclination, high-$W_0$ galaxies in
our sample.

Finally we note that we were unable to extract any further useful
information (for TFR purposes) from the rotation curves beyond the
width $W$.  A crude measure of rotation curve shape did not improve the
TFR fit, nor did we find shape to be correlated with $W$ (and
presumably galaxy mass).  While such a correlation may be present when
large numbers of galaxy rotation curves are co-added, or if a broader
range of intrinsic galaxy sizes are studied, the galaxy-to-galaxy
variations seem large enough to mask the effect for our individual
objects despite the high S/N of our rotation curves.

\begin{acknowledgements}
We thank the referee, Brent Tully, for his helpful hints.
SR was supported by a Smithsonian postdoctoral fellowship
from the Smithsonian Institution
during a significant part of this research.  GMB was supported for
much of this work by the Bok Fellowship at Steward Observatory. PG is
supported by a NASA Long Term Space Astrophysics grant NAG~5-3232.
KvB was forced to teach Astronomy 102 to support himself.
This research has made use
of the NASA/IPAC Extragalactic Database (NED), which is operated by
the Jet Propulsion Laboratory, Caltech, under contract with the NASA.
\end{acknowledgements}

\clearpage
\def\kms{{km~s$^{-1}$}}
\font\smcap=cmcsc10
\def\hi{{\smcap H$\,$i}}
\def\hii{{\smcap H$\,$ii}}
\def\ha{H$\alpha$}
\begin{deluxetable}{rcccccccccc}
\footnotesize
\tablewidth{0pt}
\tablecaption{Galaxy Data}
\tablehead{
\colhead{Code} & \colhead{Name} &
\colhead{$I'$} & \colhead{$e$} &
\colhead{PA} &\colhead{$r_{1/2}$} &
\colhead{$v$} &
\colhead{$W$ (\ha)} & 
\colhead{$W$ (\ha)} & 
\colhead{$W$ (21-cm)} & 
\colhead{$f$} 
\nl
  &  & \colhead{[mag]} &   & \colhead{[$^\circ$]} &
\colhead{[$^{\prime\prime}$]} & \colhead{[\kms]} &
\colhead{[\kms]} & \colhead{[\kms]} & \colhead{[\kms]} &  \nl
\colhead{(1)} & \colhead{(2)} & \colhead{(3)} & \colhead{(4)} & \colhead{(5)} & 
\colhead{(6)} & \colhead{(7)} & \colhead{(8)} & \colhead{(9)} & \colhead{(10)} & \colhead{(11)} }
\startdata
SC24	& 128-087	& 13.16	& 0.73	& 80	& $\ $6.5   & 6672    & 384     & 354 	& 404	&   0.73 \nl
SC25	& 158-105	& 13.03	& 0.62	& $\llap{--}$79	& $\ $3.9   & 6821    & 370     & 342 	& 393	&   0.85 \nl
SC28	& 129-010	& 13.51	& 0.74	& 27	& $\ $4.9   & 7016    & 330     & 312 	& 366	&   0.77 \nl
SC31	&  99-104	& 13.77	& 0.70	& 12	& $\ $4.7   & 7934    & 339     & 328 	& 382	&   0.74 \nl
SC32	& 159-055	& 14.07	& 0.72	& 74	& $\ $3.0   & 7750    & 285     & 234 	& 318	&   0.71 \nl
SC39	& 159-096	& 13.52	& 0.64	& $\ $8	& $\ $2.3   & 6189    & 292     & 274 	& 313	&   0.71 \nl
SC58	& 131-008	& 13.16	& 0.79	& 38	& $\ $4.0   & 5981    & 408     & 363 	& 434	&   0.71 \nl
C1	& 160-088	& 12.81	& 0.50	& 55	& $\ $9.4   & 7287    & 377     & 364 	& 391	&   0.74 \nl
C4	& 160-102	& 12.78	& 0.71	& 88	& $\ $2.7   & 7103    & 464     & 451 	& 498	&   0.80 \nl
C5	& 130-012	& 12.65	& 0.81	& $\llap{--}$38	& $\ $6.0   & 7130    & 526     & 495 	& 554	&   0.67 \nl
C9	& 160-137	& 12.17	& 0.43	& $\llap{--}$39	& $\ $2.8   & 7030    & 388     & 372 	& 459	&   0.88 \nl
C11	& 160-166	& 11.64	& 0.44	& 22	& 20.1  & 6411    & 542     & 537 	& 567	&   0.68 \nl
C12	& 160-192	& 12.07	& 0.61	& $\llap{--}$78	& 26.7  & 6648    & 518     & 509 	& 564	&   0.72 \nl
C13	& 159-059	& 13.64	& 0.30	& 24	& $\ $9.4   & 7519    & 202     & 190 	& 216	&   0.77 \nl
C14	& 159-082	& 13.21	& 0.38	& 11	& 13.7  & 8087    & 305     & 290 	& 321	&   0.73 \nl
C15	& 159-099	& 13.75	& 0.70	& $\llap{--}$86	& 17.0  & 7886    & 372     & 341 	& 392	&   0.78 \nl
C16	& 159-102	& 12.52	& 0.70	& 34	& 13.2  & 7071    & 529     & 509 	& 551	&   0.74 \nl
C17	& 159-110	& 12.48	& 0.79	& 72	& 17.2  & 6330    & 522     & 458 	& 541	&   0.56 \nl
C19	& UGC8195	& 14.47	& 0.86	& 89	& 20.8  & 7038    & 263     & 227 	& 268	&   0.65 \nl
C21	& UGC8244	& 14.07	& 0.62	& 78	& 20.8  & 7099    & 237     & 226 	& 269	&   0.72 \nl
C23	& 160-167	& 13.31	& 0.56	& $\llap{--}$43	& 15.7  & 6035    & 293     & 270 	& 308	&   0.66 \nl
C26	& 159-080	& 13.59	& 0.73	& $\ $7	& 12.3  & 6897    & 342     & 306 	& 360	&   0.74 \nl
C27	& 159-106	& 14.02	& 0.58	& $\ \llap{--}$1	& $\ $9.0   & 7954    & 301     & 270 	& 311	&   0.69 \nl
C31	& UGC7955	& 13.42	& 0.82	& 29	& 19.4  & 6752    & 372     & 344 	& 412	&   0.76 \nl
SC29	& 159-018	& 13.05	& 0.62	& $\llap{--}$39	& 12.2  & 8052    & 469     & 445 	& 451	&   0.82 \nl
\enddata
\tablecomments{
Columns~(1) and (2) give our internal name and the Zwicky or UGC name of
each galaxy,
respectively.  Column~(3) is the total $I$-band magnitude, corrected to a
distance of 7000 \kms\ as per the left-hand side of Equation~(1).
Column~(4) gives the ellipticity determined from the $I$-band surface
photometry (see Paper~I), 
column~(5) is the position angle used for the long-slit observations, and
column~(6) is the semi-major axis of the isophote enclosing half the $I$-band
light.  The recession velocity in column~(7) is determined from the optical
data.  Widths in column~(8) are from the Probable Min-Max method,
in column~(9) using the Weighted Mean method over the range 1--2$r_{1/2}$,
and in column (10) we repeat the published 21-cm width (Paper~I).  Column~(11)
gives the rotation curve shape parameter defined in Equation~(2).
}
\end{deluxetable}
\clearpage

\begin{deluxetable}{lccc}
\tablewidth{0pt}
\tablecaption{Tully-Fisher Results for Various Algorithms}
\tablehead{
 Algorithm & $m$  & $a_{I}$  & RMS scatter \nl
           & [mag] & [mag]  & [mag]
}
\startdata
 Min-Max & --5.41 $\pm$ 0.26 & 1.46 $\pm$ 0.20 & 0.140 \nl
 Weighted Mean: Entire Side & --3.56 $\pm$ 0.22 & 1.33 $\pm$ 0.25 & 0.175 \nl
 Weighted Mean: Outer Half & --5.47 $\pm$ 0.26 & 1.35 $\pm$ 0.20 & 0.138 \nl
 Weighted Mean: 1--2$\,r_{1/2}$ & --4.95 $\pm$ 0.24 & 1.18 $\pm$ 0.20 & 0.139 \nl
 Weighted Mean: 1--4$\,r_{1/2}$ & --5.08 $\pm$ 0.22 & 1.25 $\pm$ 0.18 & 0.128 \nl
 Fixed Point: 1.3$\,r_{1/2}$ & --5.00 $\pm$ 0.24 & 1.45 $\pm$ 0.20 & 0.141 \nl
 Probable Min-Max\tablenotemark{\dag} & --5.40 $\pm$ 0.25 & 1.47 $\pm$ 0.25 & 0.135  \nl
\tablevspace{8pt}
 21-cm Widths     & --5.61 $\pm$ 0.18 & 1.42 $\pm$ 0.13 & 0.094  \nl
\enddata
\tablenotetext{\dag}{This algorithm is used to produce Figures~2 and 3.}
\end{deluxetable}

\end{document}